\documentclass[conference]{IEEEtran}
\IEEEoverridecommandlockouts
\usepackage{cite}
\usepackage{amsmath,amssymb,amsfonts}
\usepackage{algorithmic}
\usepackage{graphicx}
\usepackage{textcomp}
\usepackage{xcolor}
\usepackage[switch]{lineno}
\usepackage{hyperref}
\usepackage{wrapfig}
\usepackage{url}
\usepackage[english]{babel}
\usepackage{tabu}
\usepackage{booktabs}
\usepackage{makecell}
\usepackage{multirow}
\usepackage{amsthm,thmtools,thm-restate}
\everydisplay{\footnotesize}
\usepackage[ruled,vlined,linesnumbered]{algorithm2e}
\SetKwComment{Comment}{/* }{ */}
\RestyleAlgo{ruled}

\DeclareUnicodeCharacter{2032}{\ensuremath{\prime}}
\usepackage[draft]{changes}

\def\BibTeX{{\rm B\kern-.05em{\sc i\kern-.025em b}\kern-.08em
    T\kern-.1667em\lower.7ex\hbox{E}\kern-.125emX}}
\begin{document}

\title{FFCz: \underline{F}ast \underline{F}ourier \underline{C}orrection for Spectrum-Preserving Lossy Compression of Scientific Data
}

\author{\IEEEauthorblockN{Congrong Ren}
\IEEEauthorblockA{
\textit{The Ohio State University}\\
Columbus, Ohio \\
ren.452@osu.edu}
\and
\IEEEauthorblockN{Robert Underwood}
\IEEEauthorblockA{
\textit{Argonne National Laboratory}\\
Lemont, Illinois \\
runderwood@anl.gov}
\and
\IEEEauthorblockN{Sheng Di}
\IEEEauthorblockA{
\textit{Argonne National Laboratory}\\
Lemont, Illinois \\
sdi1@anl.gov}
\and
\IEEEauthorblockN{Emrecan Kutay}
\IEEEauthorblockA{
\textit{The Ohio State University}\\
Columbus, Ohio \\
kutay.5@osu.edu}
\and
\IEEEauthorblockN{Zarija Luki\'{c}}
\IEEEauthorblockA{
\textit{Lawrence Berkeley National Laboratory}\\
Berkeley, California \\
zarija@lbl.gov}
\and
\IEEEauthorblockN{Aylin Yener}
\IEEEauthorblockA{
\textit{The Ohio State University}\\
Columbus, Ohio \\
yener@ece.osu.edu}
\and
\IEEEauthorblockN{Franck Cappello}
\IEEEauthorblockA{
\textit{Argonne National Laboratory}\\
Lemont, Illinois \\
cappello@mcs.anl.gov}
\and
\IEEEauthorblockN{Hanqi Guo}
\IEEEauthorblockA{
\textit{The Ohio State University}\\
Columbus, Ohio \\
guo.2154@osu.edu}
}

\maketitle

\begin{abstract}
This paper introduces a novel technique to preserve spectral features in lossy compression based on a novel fast Fourier correction algorithm for regular-grid data. Preserving both spatial and frequency representations of data is crucial for applications such as cosmology, turbulent combustion, and X-ray diffraction, where spatial and frequency views provide complementary scientific insights. In particular, many analysis tasks rely on frequency-domain representations to capture key features, including the power spectrum of cosmology simulations, the turbulent energy spectrum in combustion, and diffraction patterns in reciprocal space for ptychography. However, existing compression methods guarantee accuracy only in the spatial domain while disregarding the frequency domain. To address this limitation, we propose an algorithm that corrects the errors produced by off-the-shelf ``base'' compressors such as SZ3, ZFP, and SPERR, thereby preserving both spatial and frequency representations by bounding errors in both domains. By expressing frequency-domain errors as linear combinations of spatial-domain errors, we derive a region that jointly bounds errors in both domains. Given as input the spatial errors from a base compressor and user-defined error bounds in the spatial and frequency domains, we iteratively project the spatial error vector onto the regions defined by the spatial and frequency constraints until it lies within their intersection. We further accelerate the algorithm using GPU parallelism to achieve practical performance. We validate our approach with datasets from cosmology simulations, X-ray diffraction, combustion simulation, and electroencephalography demonstrating its effectiveness in preserving critical scientific information in both spatial and frequency domains. An open source GPU implementation of our algorithm is available at \url{https://github.com/rcrcarissa/FFCz}.
\end{abstract}


\section{Introduction}

Error-bounded lossy compression has become a widely used data reduction strategy in modern scientific data analysis to maintain performance and scalability in the face of the growing gap between data generation rates and both of available storage and communication bandwidth. For example, the Nyx cosmology code produces 2.8 PB of data in just five simulation runs, each consisting of 200 snapshots~\cite{almgren2013nyx}, placing significant pressure on storage systems and data movement. Unlike lossless compression that reduces data without information loss yet achieves limited compression ratios (around 2:1 for floating-point data~\cite{zhao2021optimizing}), error-bounded lossy compression achieves substantially higher ratios while maintaining data quality, and has been successfully applied across a wide range of scientific domains, including cosmology~\cite{zhang2025lcp}, fluid dynamics~\cite{zou2020performance}, and medical studies~\cite{chen2012dynamic} and widely adopted to reduce storage costs~\cite{ren2024prediction,wu2025enabling}, accelerate I/O performance~\cite{liang2019improving}, and facilitate large-scale data analysis~\cite{di2025survey,liang2020toward}.

\begin{wrapfigure}{LH}{0.5\linewidth}
    \centering
    \includegraphics[width=\linewidth]{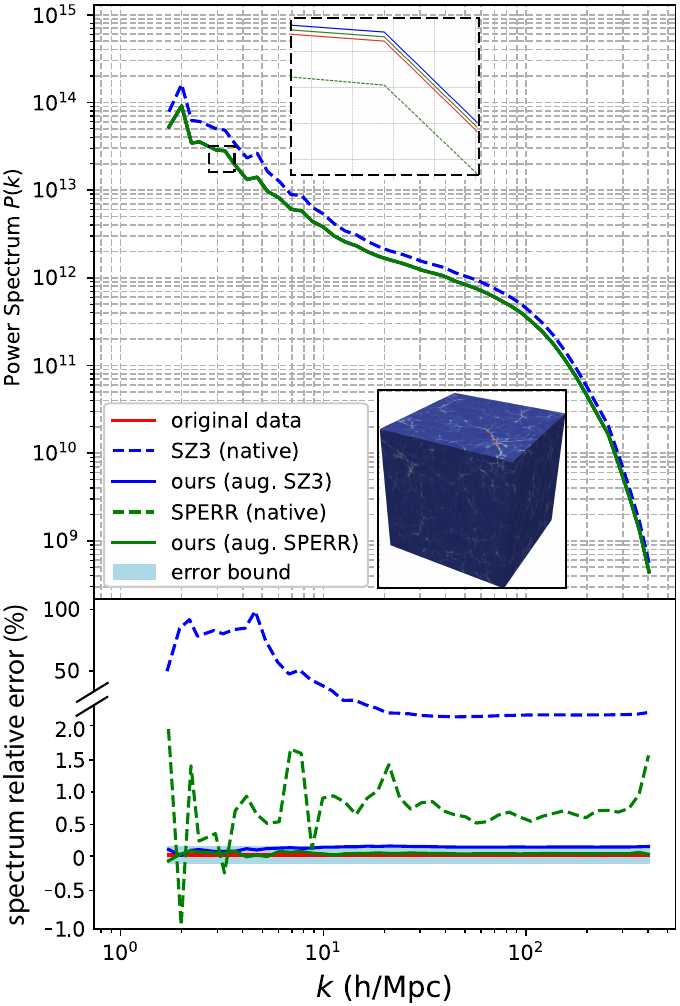}
    \caption{Power spectra of the baryon density field from the Nyx $512^3$ dataset, compressed with SZ3 and SPERR and edited by our method, at the same bitrate (i.e., the number of bits used to encode a single value) of 0.023. Spatial and spectral relative error bounds are $0.01\%$ and $0.1\%$, respectively.}
    \label{fig:teaser}
\end{wrapfigure}

A key limitation of existing error-bounded lossy compression algorithms is that they are primarily designed to bound reconstruction error in the original domain (typically spatial or temporal; referred to as the spatial domain for brevity throughout this paper), overlooking distortions in the frequency domain, which is central to many data-driven applications. For example, the power spectrum of cosmology data, key to analyzing matter and energy distribution across spatial scales, can be severely distorted if only spatial errors are bounded while frequency components are inaccurately reconstructed, as showcased in Fig.~\ref{fig:teaser}. Two-point correlation analysis of turbulence simulation data relies on preserving energy across scales, and errors in the frequency domain can distort estimates of energy spectra~\cite{mouri2017two}. In electroencephalography data, failure to preserve the frequency domain can lead to misinterpretation of neural rhythms and compromise detection of clinically relevant patterns~\cite{li2021fft}.

Despite the importance of spectral fidelity, existing compressors lack a direct and rigorous mechanism to control frequency-domain accuracy. To date, preservation of spectral features has been done in an indirect and empirical manner. For example, users must manually tune compression parameters (e.g., spatial error bounds) through repeated trial and error until the desired spectral accuracy is achieved. This procedure not only requires substantial effort to empirically identify feasible spatial error bounds but also forces the compressor to satisfy overly strict spatial bounds, significantly reducing the achievable compression ratio. These challenges highlight the need for novel approaches that can balance compression ratio, dual-domain accuracy, and performance on large datasets.

We propose a dual-domain error bounding paradigm with an end-to-end GPU implementation for lossy compression to preserve data fidelity in both the spatial and frequency domains, using an edit-based strategy applicable to datasets in 1D, 2D, 3D, and beyond. To the best of our knowledge, this is the first algorithm enabling users to rigorously enforce error bounds simultaneously in both the dual domains. Our core idea is to model frequency-domain errors as linear combinations of spatial-domain errors with complex-valued coefficients, allowing us to characterize and confine the space of spatial-domain errors that satisfy constraints in both domains. Specifically, we define the \textit{feasible region} for spatial-domain errors as the intersection of two geometric structures: an axis-aligned (hyper) cube (or \textit{$s$-cube} for short) derived from the spatial error bounds, and a rotated (hyper)cube (or \textit{$f$-cube} for short) induced by the frequency-domain error bounds. Given compressed data from an arbitrary lossy compressor (so-called the \textit{``base'' compressor}, e.g., SZ3~\cite{liang2022sz3}) that satisfies the spatial error constraints, we first identify frequency components whose errors exceed the frequency error bounds and project the spatial error vector onto the $f$-cube. Next, we locate spatial components that violate the spatial error constraint and project the error vector onto the $s$-cube. We iteratively repeat these projections onto the two cubes until the error vector lies within the intersection of the two cubes, thereby adjusting the data to meet the requirements of both domains. The resulting edits are then given by the adjustments to the spatial error vector required to enter the feasible region. We further quantize and compress the edits to reduce storage overhead (up to 10\% of the compressed storage to store the edits as shown later). To further improve efficiency, we exploit GPU parallelism for projecting a point onto a cube and for verifying constraint satisfaction across data blocks.

We evaluate our method through a comprehensive comparison with state-of-the-art error-bounded lossy compressors. Specifically, we use SZ3, ZFP~\cite{lindstrom2006fast}, and SPERR~\cite{li2023lossy} as base compressors. Experimental results demonstrate that our approach consistently achieves better frequency-domain preservation and comparable spatial-domain preservation under the same bitrate, along with fast performance across diverse datasets from domains such as cosmology, ptychography, and combustion. In addition, we assess the accuracy on the \textit{power spectrum}, a critical metric for characterizing the statistical distribution of matter or fluctuations in many physical simulations. We make the following contributions:
\begin{itemize}
    \item A novel algorithm for preserving information in both spatial and frequency domains of multi-dimensional datasets, enabling dual-domain accuracy guarantees;
    \item An efficient GPU-parallel algorithm that accelerates the proposed correction algorithm;
    \item Adoption of new metrics for frequency domain, including spectral signal-to-noise ratio (SSNR) and relative error on power spectrum;
    \item A comprehensive evaluation across diverse real-world datasets and three state-of-the-art compressors (SZ3, ZFP, and SPERR).
\end{itemize}
\section{Related Work}

We summarize the relevant literature on error-bounded lossy compression, review the Fourier transform and its applications, and discuss its connection to spectral-domain analysis.

\subsection{Error-bounded lossy compression}

Error-bounded lossy compressors ensure that each reconstructed data point deviates from its original value by no more than a user-specified bound, thereby controlling errors in the spatial domain, but they provide no guarantees on deviations in the frequency domain. For example, transform-based methods such as ZFP~\cite{lindstrom2014fixed} and SPERR~\cite{li2023lossy} apply blockwise orthogonal transforms (e.g., integer-to-integer or wavelet-like bases), but these transforms are local rather than spectral, capturing neighboring correlation rather than global frequency components. Prediction-based compressors, such as SZ~\cite{liang2018error,liang2022sz3,tao2017significantly,zhao2021optimizing,zhao2020significantly}, perform pointwise prediction and quantization, ensuring bounded deviations per data point but without accounting for correlations across frequencies. As a result, even with tightly controlled spatial errors, the cumulative effect in the frequency domain can be unbounded, causing distortions in power spectra and other frequency-dependent analyses, as shown in evaluation section.

\subsection{Feature-preserving lossy compression}

Researchers have made efforts to preserve quantities of interest (QoIs) rather than only spatial-domain accuracy in lossy compression, shifting the focus from pointwise fidelity toward scientifically meaningful properties that support post-hoc analyses. These QoIs include topological, statistical, or physical invariants~\cite{di2025survey}. For example, Liu et al. proposed QPET~\cite{liu2024qpet}, which preserves the quantile distribution of data to maintain key statistical characteristics.

Building on this direction, post-compression augmentation- or edit-based paradigms have emerged as effective methods for enforcing QoI constraints. These methods operate on the decompressed output of a base compressor, introducing \textit{edits} to the decompressed output to enforce QoI constraints without violating the original error bound. For example, Li et al. introduced MSz, which derives edits that preserve a topological descriptor named Morse–Smale segmentations in 2D and 3D piecewise linear scalar fields~\cite{li2024msz}. Gorski et al. developed a framework that quantifies and applies adjustments to preserve the contour tree~\cite{gorski2025general}. However, preserving spectral features is still an open problem, as further discussed below.

\subsection{Fourier and spectral domain analysis}

Fourier transform-based methods enable efficient feature extraction and pattern recognition in large-scale data analysis. \textit{Fast Fourier transform} (FFT), an accelerated Fourier transform calculation algorithm, also accelerates fundamental signal processing operations such as convolution, filtering, and correlation, which are commonly used for feature extraction in large-scale data environments~\cite{bulgakov2024dimensionality,buneman1993fast,marten2019calculating}. For instance, in medical datasets, Fourier transform is widely used to classify signal segments into categories for detecting medical conditions and analyzing mental tasks~\cite{li2021fft}. In multimedia datasets, it supports speech, music, and noise classification in audios~\cite{chu2007noise}, similarity search for images~\cite{celentano1997fft,tsai2012fft}, and moving object detection in videos~\cite{asadzadehkaljahi2025spatio}. 

The \textit{power spectrum}, a particularly important Fourier-derived quantity, reveals the power distribution across different frequency components~\cite{van2011effects}. The shape of the power spectrum curve, characterized by the wavenumber $k$ and power spectrum $P(k)$, provides plenty of insight into the underlying structure of the data. Peaks in the spectrum correspond to dominant periodic components, while flat regions indicate white noise with uniformly distributed power. Many physical processes follow a \textit{power-law} behavior, $P(k)\propto k^{-\alpha}$, visible as a straight line on a log–log plot, where $\alpha$ measures the relative strength of large- and small-scale fluctuations~\cite{burgess2007signal}. In contrast, an \textit{exponential decay}, $P(k)\propto e^{-k/k_0}$, indicates strong suppression of high-frequency content, typical of smooth fields~\cite{sigeti1995survival}. While Jin et al. performed fine-grained rate–quality modeling to configure existing compressors~\cite{jin2021adaptive}, our work proposes a novel algorithm explicitly designed to preserve Fourier-domain fidelity, including power spectrum accuracy.

\section{Background}

\textbf{Discrete Fourier transforms}.
\textit{Discrete Fourier transform} (DFT)~\cite{sundararajan2001discrete} converts a discrete signal from the spatial domain to the frequency domain. For a 1D dataset with $N$ data points denoted by $\{x_n\}_{n=0}^{N-1}$, DFT calculates its frequency components $X_k$ by
\begin{equation}
    X_k=\sum_{n=0}^{N-1}x_n e^{-i\frac{2\pi k}{N}n},\ k\in\{0,\ 1,\ ...,\ N-1\},
\label{eq:dft}
\end{equation}
For a 2D dataset $\{x_{m,n}\}$ with dimensionality of $N_0N_1$ and its frequency components $\{X_{u,v}\}$, the DFT generalizes as
\begin{equation*}
\begin{aligned}
    X_{u,v}&=\sum_{m=0}^{N_0-1}\sum_{n=0}^{N_1-1}x_{m,n}e^{-i2\pi\left(\frac{um}{N_0}+\frac{vn}{N_1}\right)},\\
    &u\in\{0,\ 1,\ ...,\ N_0-1\},\ v\in\{0,\ 1,\ ...,\ N_1-1\}
\end{aligned}
\end{equation*}
Higher-dimensional DFTs follow the same pattern with additional exponential terms per dimension.

\textbf{Power spectrum}. The power spectrum (denoted by $P(k)$) is a fundamental tool for analyzing signals, enabling the detection of dominant frequencies and the identification of patterns such as white noise, which appears as a flat spectrum. Given a discrete 3D dataset with data values $\{x_{m,n,p}\}$, we first normalize the fluctuations to remove the mean background by
\begin{equation*}
    x'_{m,n,p}=\frac{x_{m,n,p}-\bar{x}}{\bar{x}},
\end{equation*}
where $\bar{x}$ is the average value over $\{x_{m,n,p}\}$. This step ensures the power spectrum measures relative structure and variability instead of the absolute field magnitude. Then we shift the zero-frequency component of $\{x'_{m,n,p}\}$ to the center to make clear the symmetry between positive and negative frequencies~\cite{zero-shift}. Next, the spatial function is transferred into the frequency domain:
\begin{equation*}
    X'_{u,v,w}=\mbox{FFT}(x'_{m,n,p}).
\end{equation*}
The power spectrum then accumulates the magnitudes of frequencies that have the same distance from the center:
\begin{equation*}
    P(k)=\sum_{u^2+v^2+w^2=k^2}|X'_{u,v,w}|^2,
\end{equation*}
where $|\cdot|$ means taking the magnitude of a complex value.

\begin{figure}[!th]
    \centering
    \includegraphics[width=.8\linewidth]{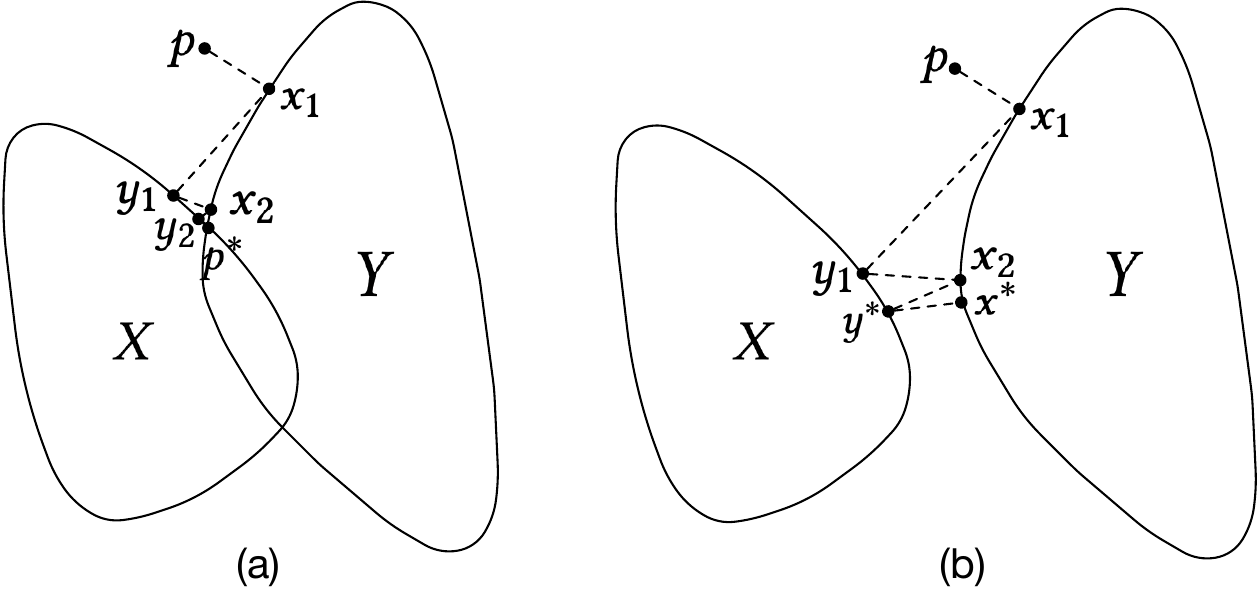}
    \caption{POCS method. It iteratively projects an arbitrary point back and forth between two closed convex sets until reaching a point in their intersection. The convergence is guaranteed whenever the two sets have a non-empty intersection.}
    \label{fig:alternating_projection}
\end{figure}

\textbf{Projections onto Convex Sets (POCS) Method}.
Our algorithm that moves a error vector into the intersection of spatial and frequency constraints is a specialized adaptation of the Projections onto Convex Sets (POCS) method, a well-established iterative technique for finding a point in the intersection of two closed convex shapes. As shown in Fig.~\ref{fig:alternating_projection}~{(a)}, given two (closed) shapes $X$ and $Y$, the procedure begins with an arbitrary initial point $p$, and repeatedly projects it alternately onto each set. When the intersection $X\cap Y$ is nonempty, the iterates are guaranteed to converge to a point $p^*\in X\cap Y$. In contrast, if $X$ and $Y$ do not intersect, the iterates cannot converge to a common point and instead oscillate between two points $x^*$ and $y^*$ lying on the respective sets (Fig.~\ref{fig:alternating_projection}~{(b)}). When finding the point in the intersection of two hypercubes, the alternating projection algorithm exhibits linear (geometric) convergence toward a point in their intersection. The convergence rate depends on the relative orientation of the hypercubes and the set of active faces at the intersection. When the intersection is nearly tangential, convergence is slow; conversely, a more transversal intersection yields faster convergence.

Although alternative methods for projecting onto the intersection of convex sets could be used in our workflow, we adopt the POCS method due to its simplicity, memory efficiency, and suitability for parallel implementation on GPUs. Methods such as Dykstra’s algorithm and the Alternating Direction Method of Multipliers (ADMM) often converge faster, but they incur higher memory costs for storing correction terms or primal–dual variables. Similarly, Douglas–Rachford splitting can exhibit rapid convergence; however, its iterates may converge to points outside the intersection of the two sets.

\section{Methodology}
\label{sec:method}

This section first formulates the problem of error-bounding data in both the spatial and frequency domains (so called ``dual domains''), then describes our algorithm using 1D data as an example, generalizes it to multi-dimensional data, and introduces its parallel GPU implementation.

\begin{figure*}
    \centering
    \includegraphics[width=\textwidth]{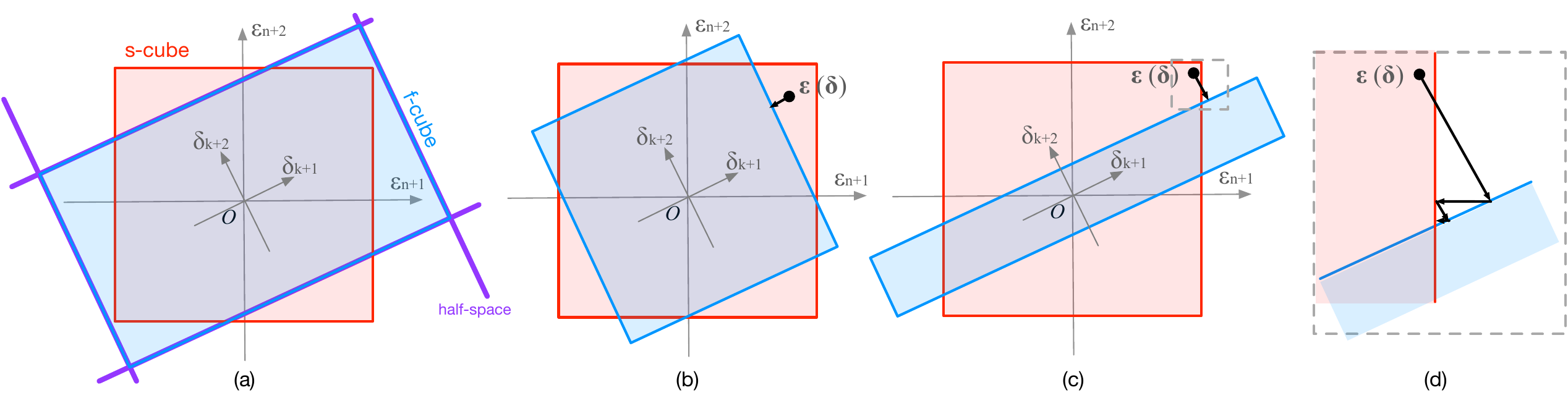}
    \caption{Illustration of our method. (a) Feasible region for spatial error vector $\pmb{\epsilon}$, where $\epsilon_n$ and $\delta_k$ denote spatial and frequency bases. Red and blue rectangles represent $s$-cube and $f$-cube constraints. The half-space pair along $\delta_{k+2}$ is redundant, while that along $\delta_{k+1}$ is not. (b) A spatial error vector inside the $s$-cube but outside the $f$-cube is projected onto the $f$-cube by clipping its frequency error vector $\pmb{\delta}$. If it remains within the $s$-cube, the process stops. (c) Otherwise, it is iteratively projected between the $s$-cube and $f$-cube until reaching their intersection. (d) Detailed view of the alternating projections.}
    \label{fig:method}
\end{figure*}

\subsection{Dual-domain bounding problem}

The dual-domain bounding problem is formulated to bound the error of reconstructed data in both spatial and frequency domains. Given the original data $\{x_n\}_{n=0}^{N-1}$ and the decompressed data $\{\hat{x}_n\}_{n=0}^{N-1}$ from a base lossy compressor, we denote the $n$-th spatial-domain error as $\epsilon_n=\hat{x}_n-x_n$ and the $k$-th frequency-domain error as $\delta_k=\hat{X}_k-X_k$, where $X_k$ and $\hat{X}_k$ are the $k$-th frequency components obtained by applying the FFT to $\{x_n\}_{n=0}^{N-1}$ and $\{\hat{x}_n\}_{n=0}^{N-1}$, respectively. Without loss of generality, we describe our algorithm with user-specified global spatial error bounds $E$ and frequency error bound $\Delta$:\footnote{We will talk about the generalization of the error bounds to pointwise fashion $E_n$, $\Delta_k^\Re$, and $\Delta_k^\Im$ in later sections.}
\begin{equation}
    \begin{aligned}
        &|\epsilon_n|\leq E,\\
        &|\Re(\delta_k)|\leq\Delta\mbox{ and }|\Im(\delta_k)|\leq\Delta,
    \end{aligned}
\label{eq:constraints_v1}
\end{equation}
for all $n$ and $k$ in $\{0,\ 1,\ ...,\ N-1\}$, where $|\cdot|$, $\Re(\cdot)$, and $\Im(\cdot)$ mean taking absolute value, real part, and imaginary part.

\textbf{Mapping frequency-domain constraints to spatial domain}.  Based on the definition of DFT (Eq.~\eqref{eq:dft}), the frequency-domain error $\delta_k$ is a linear combination (with complex weights) of all spatial-domain errors $\epsilon_n$:
\begin{equation}
\delta_k=\hat{X}_k-X_k=\sum_{n=0}^{N-1}\epsilon_n\exp\left(-i\frac{2\pi k}{N}n\right),\ k\in\{0,\ 1,\ ...,\ N-1\},
\label{eq:frequency_err}
\end{equation}
Thus, one can express the constraints in Eq. ~\eqref{eq:constraints_v1} that include both spatial errors $\epsilon_n$ and frequency errors $\delta_k$ by the following form that only contains spatial errors $\epsilon_n$:
\begin{subequations}\label{eq:constraints_v2}
\begin{align}
    -\Delta\leq\sum_{n=0}^{N-1}\cos&\left(\frac{2\pi k}{N}n\right)\cdot\epsilon_n\leq\Delta, \label{eq:constraints_v2a}\\
    -\Delta\leq\sum_{n=0}^{N-1}\sin&\left(\frac{2\pi k}{N}n\right)\cdot\epsilon_n\leq\Delta, \label{eq:constraints_v2b}\\
    -E\leq&\epsilon_n\leq E\label{eq:constraints_v2c},
\end{align}
\end{subequations}
for all $n$ and $k$ in $\{0,\ 1,\ ...,\ N-1\}$.

\textbf{$f$-cube and $s$-cube}. We refer to the constraints in Eq.~\eqref{eq:constraints_v2a} and~\eqref{eq:constraints_v2b}, which bound the errors in the frequency domain, as the \textit{$f$-cube}, and the constraint in Eq.~\eqref{eq:constraints_v2c}, which bounds errors in the spatial domain, as the \textit{$s$-cube}, based on geometric shapes induced in the space of \textit{spatial error vector} $\pmb{\epsilon}=\{\epsilon_0, \epsilon_1,...,\epsilon_{N-1}\}\in\mathbb{R}^N$ (Fig.~\ref{fig:method}~{(a)}). Both of $s$- and $f$-cubes are with the dimensionality of the number of data points. For convenience, we call any constraint on $\pmb{\epsilon}$ of the form $\mathbf{a}_k^\intercal\pmb{\epsilon}\leq\Delta$ for some vector $\mathbf{a}_k\in\mathbb{R}^N$ associated with frequency component $k$ a \textit{half-space constraint}. A constraint of the form $-\Delta\leq\mathbf{a}_k^\intercal\pmb{\epsilon}\leq\Delta$ is then a \textit{pair of half-space constraints}. The $n$-th element of $\mathbf{a}_k$, denoted $a_{k,n}$, is given by $a_{k,n}=\cos{(\frac{2\pi k}{N}n)}$ for the real part of frequency component $k$, or $a_{k,n}=\sin{(\frac{2\pi k}{N}n)}$ for the imaginary part. The $f$-cube is thus the intersection of such pairs of half-space constraints over a set of frequency components.

\begin{figure*}
    \centering
    \includegraphics[width=\textwidth]{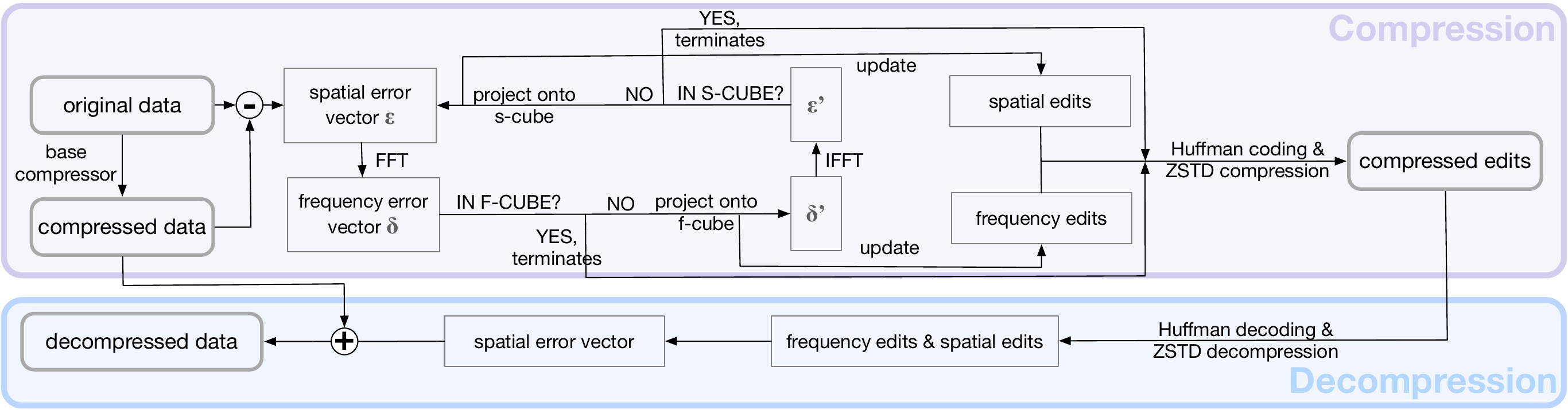}
    \caption{Pipeline of our alternating projection algorithm. We iteratively project the error vector onto the $f$- and $s$-cubes until it converges to a point in their intersection.}
    \label{fig:workflow}
\end{figure*}

\textbf{Frequency and spatial bases}. Consider the collection of $2N$ of cosine and sine vectors $\{\mathbf{a}_k\}$ that defines the normal vectors of half-space constraints for $k\in\{0,1,...,N-1\}$. By Hermitian symmetry~\cite{bienenstock1962symmetry}, i.e., $X_{N-k}=X^*_k$ where $X^*$ denotes complex conjugation, only frequency indices $k\leq N/2$ contribute distinct normal vectors. Furthermore, for $k=0$ or $k=N/2$, the sine term vanishes, leaving only the cosine component. Altogether, the $2N$ vectors $\{\mathbf{a}_k\}$ yield only $N$ distinct vectors. These $N$ distinct vectors are mutually orthogonal under the Euclidean inner product~{stein2011fourier}, forming an alternative orthonormal \textit{frequency basis} for $\mathbb{R}^N$, distinct from the standard \textit{spatial basis} defined by $\pmb{\epsilon}$. In this representation, the $f$-cube is axis-aligned in the frequency basis (rotated in the spatial basis), while the $s$-cube is axis-aligned in the spatial basis (rotated in the frequency basis), as illustrated in Fig.~\ref{fig:method}~{(a)}.

\subsection{Alternating Projection-Correction Algorithm}
\label{sec:workflow}

Fig.~\ref{fig:workflow} and Alg.~\ref{alg:alternating_projection} show the workflow and pseudo-code of our alternating projection-correction algorithm, respectively. Given original data $\{x_n\}_{n=0}^{N-1}$ and decompressed data $\{\hat{x}_n\}_{n=0}^{N-1}$ from a base compressor, the algorithm iteratively projects the reconstruction error onto the $f$- and $s$-cubes, recording the displacement of the error vector as \textit{spatial edits} along the spatial basis and \textit{frequency edits} along the frequency basis.

\begin{algorithm}[!ht]
\caption{Alternating Projection-Correction}\label{alg:alternating_projection}
\scriptsize
\KwData{original data $\mathbf{x}$, decompressed data $\mathbf{\hat{x}}$ from a base compressor;\\
\hspace{0.85cm}$E$: user-defined spatial error bound;\\
\hspace{0.85cm}$\Delta$: user-defined frequency error bound}
$\pmb{\epsilon}$ $\gets$ $\mathbf{\hat{x}}-\mathbf{x}$\;
spat\_edits $\gets$ []\Comment*[r]{spatial edits}
freq\_edits $\gets$ []\Comment*[r]{frequency edits}
\While{\texttt{True}}{
$\pmb{\delta}$ $\gets$ \texttt{FFT}($\pmb{\epsilon}$)\;
\If{\texttt{CheckConvergence($\pmb{\delta}$,$\Delta$)}}{\texttt{break}}
$\pmb{\delta}'$ $\gets$ \texttt{ProjectOntoFCube}($\pmb{\delta}$, $\Delta$)\;
freq\_edits $+=\pmb{\delta}'-\pmb{\delta}$\;
$\pmb{\delta}$ $\gets$ $\pmb{\delta}'$\;
$\pmb{\epsilon}$ $\gets$ \texttt{IFFT}($\pmb{\delta}$)\;
$\pmb{\epsilon}'$ $\gets$ \texttt{ProjectOntoSCube}($\pmb{\epsilon}$, $E$)\;
spat\_edits $+=\pmb{\epsilon}'-\pmb{\epsilon}$\;
$\pmb{\epsilon}$ $\gets$ $\pmb{\epsilon}'$\;
}
spat\_flags, compact\_spat\_edits $\gets$ \texttt{CompactEdits}(spat\_edits)\;
freq\_flags, compact\_freq\_edits $\gets$ \texttt{CompactEdits}(freq\_edits)\;
compact\_quant\_spat\_edits $\gets$ \texttt{QuantizeEdits}(compact\_spat\_edits)\;
compact\_quant\_freq\_edits $\gets$ \texttt{QuantizeEdits}(compact\_freq\_edits)\;
compressed\_spat\_edits $\gets$ \texttt{LosslesslyCompressEdits}(compact\_quant\_spat\_edits)\;
compressed\_freq\_edits $\gets$ \texttt{LosslesslyCompressEdits}(compact\_quant\_freq\_edits)\;
\textbf{return} spat\_flags, freq\_flags, compressed\_spat\_edits, compressed\_freq\_edits
\end{algorithm}

\textbf{Alternating projection with $f$- and $s$- cubes}. Starting from the spatial error vector $\pmb{\epsilon}$ of the decompressed data, which initially lies within the $s$-cube, we iteratively project it onto the $f$- and $s$-cubes using a POCS-like approach, until it reaches the feasible region defined by the intersection of $f$- and $s$-cubes. We first obtain the \textit{frequency error vector} $\pmb{\delta}=\{\delta_0,\delta_1,...,\delta_{N-1}\}$ by performing an FFT on $\pmb{\epsilon}$ (as shown in Eq. ~\eqref{eq:frequency_err}), which is equivalent to the coordinate transformation from the spatial basis to the frequency basis. Violated half-space constraints are detected by comparing all elements in $\pmb{\delta}$ with $\pm\Delta$. If none are violated, then current $\pmb{\epsilon}$ already satisfies the $f$-cube constraint, and the algorithm terminates. Otherwise, we project $\pmb{\epsilon}$ onto the $f$-cube, which can be achieved by clipping $\pmb{\delta}$ to satisfy the bound $\pm\Delta$. Next, we perform an inverse FFT (IFFT) on the updated $\pmb{\delta}$ to transform back to the spatial basis, yielding an updated $\pmb{\epsilon}$, and project the updated $\pmb{\epsilon}$ onto the $s$-cube by clipping the coordinates that exceed the bound $\pm E$. This process, alternating projection onto the $f$- and $s$-cube, is repeated until $\pmb{\epsilon}$ lies within both constraints (Fig.~\ref{fig:method} (d)). Note that both $\pmb{\epsilon}$ and $\pmb{\delta}$ represent the same error vector but in different bases.

\textbf{Compaction, quantization, and lossless compression of edits}. During alternating projection, we store the shifts of the error vector $\pmb{\epsilon}$ along the spatial and frequency bases separately as \textit{spatial edits} and \textit{frequency edits}. These sequences are then converted into a compact form to reduce the storage overhead introduced on top of the base compressor. After the projection terminates, each edit sequence is further decomposed into two vectors, \textit{flags} and \textit{compact edits}. The flags are binary vectors of length $N$ indicating whether the corresponding spatial or frequency component has non-zero value; these are packed into 8-bit integers for compact storage. The compact edits store only the nonzero entries, which are typically far fewer than $N$. Compact edits are quantized by dividing each axis of the $s$-cube or $f$-cube into $2^m$ intervals, where $m$ denotes the quantization code length in bits. We fix $m=16$ for all experiments, which provides sufficient precision without increasing storage overhead. To ensure that the quantized edits still move $\pmb{\epsilon}$ into the feasible region, we slightly shrink the initial error bounds $E$ and $\Delta$ to $E(1-2^{-m})$ and $\Delta(1-2^{-m})$, respectively. Spatial and frequency edits are stored separately, since a frequency edit influences all spatial edits after inverse transform, producing a dense set of nonzero entries and thus becoming memory-inefficient. Finally, both flags and quantized compact edits are compressed using Huffman coding~\cite{huffman1952method} followed by ZSTD~\cite{zstd} to further reduce storage requirements.

\textbf{Applying edits to the decompressed data}. To reconstruct decompressed data, we first apply ZSTD decompression followed by Huffman decoding to retrieve the flags and quantized edits in both the spatial and frequency domains. Next, we dequantize these edits and reconstruct the full-length spatial and frequency edits. The \textit{complete edits} in the spatial domain are obtained by summing the spatial edits with the inverse FFT of the frequency edits. Conversely, the complete edits in the frequency domain are recovered by summing the frequency edits with the FFT of the spatial edits. Our final decompressed data is obtained by adding the complete edits to the output of the base compressor.

\subsection{Generalization to 2D and 3D problems}

Our method directly generalizes to 2D and 3D datasets without changing the formulation of $s$- and $f$-cubes. For 2D data, spatial errors $\{\epsilon_{m,n}\}$ must satisfy linear combinations corresponding to frequency bounds:
\begin{equation*}
\begin{aligned}
    -\Delta\leq&\sum_{m=0}^{N_0-1}\sum_{n=0}^{N_1-1}\cos\left(2\pi\left(\frac{um}{N_0}+\frac{vn}{N_1}\right)\right)\cdot\epsilon_{m,n}\leq\Delta\mbox{ and}\\
    -\Delta\leq&\sum_{m=0}^{N_0-1}\sum_{n=0}^{N_1-1}\sin\left(2\pi\left(\frac{um}{N_0}+\frac{vn}{N_1}\right)\right)\cdot\epsilon_{m,n}\leq\Delta\\
\end{aligned}
\end{equation*}
for all frequency indices $(u,v)$, along with spatial bounds $-E\le \epsilon_{m,n}\le E$ for all $(m,n)$. As in the 1D case, these constraints form the intersection of an $f$-cube (frequency) and an $s$-cube (spatial) in a space of dimension $N_0N_1$, with 3D data extending analogously.

\subsection{GPU Parallelism}

We parallelize our algorithm across the spatial data points and frequency components. The GPU memory hosts four classes of variables: the spatial error vector, the frequency error vector, the spatial edits and their variants (such as spatial flags and compact or quantized edits), and the frequency edits and their corresponding variants.

\textbf{Alternating projection with $f$- and $s$- cubes} (lines 5-14 in Alg.~\ref{alg:alternating_projection}). We first perform a forward FFT on the spatial error vector to obtain the frequency error vector using the highly optimized cuFFT library~\cite{cufft} (line 5). A CUDA kernel, \texttt{CheckConvergence}, then checks whether all frequency error components lie within the $f$-cube, assigning one thread per frequency component (line 6). Since the $f$-cube’s boundary hyperplanes are pairwise orthogonal, its projection operation can be decomposed into independent projections along each axis in frequency basis. If the entire frequency error vector is already within the $f$-cube, the algorithm terminates with zero spatial and frequency edits (line 7). Otherwise, we invoke the \texttt{ProjectOntoFCube} kernel to project the frequency error vector onto the $f$-cube by clipping each frequency error component to the interval $[-\Delta, \Delta]$ (line 8), update the frequency edits by the displacement along the frequency basis (line 9), and update the frequency error vector accordingly (line 10). Although these steps are implemented within a single kernel, \texttt{ProjectOntoFCube}, we list them separately in Alg.~\ref{alg:alternating_projection} to clarify how the edits and error vectors are updated. Next, an inverse FFT (via cuFFT) transforms the updated frequency error vector back to the spatial domain (line 11), followed by the \texttt{ProjectOntoSCube} kernel, which projects the spatial error vector onto the $s$-cube and updates the spatial edits and error vector with assigning one thread per spatial error (lines 12-14). Similarly, the projection onto the $s$-cube can be decomposed into independent projections along each axis in the spatial basis.

\textbf{Compaction, quantization, and lossless compression of edits} (lines 15-20 in Alg.~\ref{alg:alternating_projection}). As mentioned in Sec.~\ref{sec:workflow}, to efficiently represent the edits, we separate them into two components: binary flags indicating the positions of nonzero entries and a compact edit vector containing only the nonzero values. The exclusive prefix sum is computed in the \texttt{CompactEdits} kernel (lines 15–16) to determine the output positions of nonzero edits. Each edit in the compact vector is then quantized in parallel, with one thread assigned per edit, using the \texttt{QuantizeEdits} kernel (lines 17–18). Finally, Huffman coding is applied to losslessly compress the quantized edits. The \texttt{LosslesslyCompressEdits} kernel first divides symbols into blocks, computes symbol occurrences blockwise in parallel, and then accumulates the results to construct the global frequency table (lines 19–20).

\section{Evaluation}
\label{sec:eval}

We present the evaluation scheme and highlight key observations in this section.

\subsection{Evaluation scheme}

\textbf{Base compressors}. We select SZ3~\cite{liang2022sz3}, ZFP~\cite{lindstrom2006fast}, and SPERR~\cite{li2023lossy} as representative base compressors due to their wide usage and coverage of algorithmic designs. They are state-of-the-art and widely used lossy compressors that provide fine-grained pointwise error control across a range of scientific applications.

\textbf{Baselines}. We also use SZ3, ZFP, and SPERR as baseline methods in our comparisons, so that performance gains attributed to our approach can be directly assessed relative to the unmodified compressors.

\begin{figure*}[!ht]
    \centering
    \includegraphics[width=\textwidth]{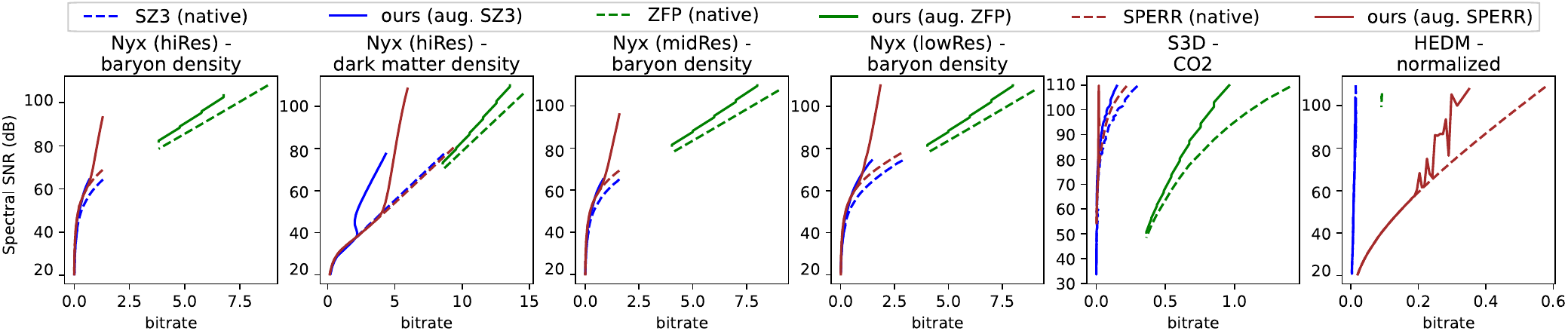}
    \caption{SSNR vs. bitrate of our method, SZ3, ZFP, and SPERR on the frequency domain. When applying our method, we edit on the results of base compressors with spatial error bound $\epsilon(\%)=0.1$.}
    \label{fig:rate_distortion}
\end{figure*}

\begin{figure*}[!ht]
    \centering
    \includegraphics[width=\textwidth]{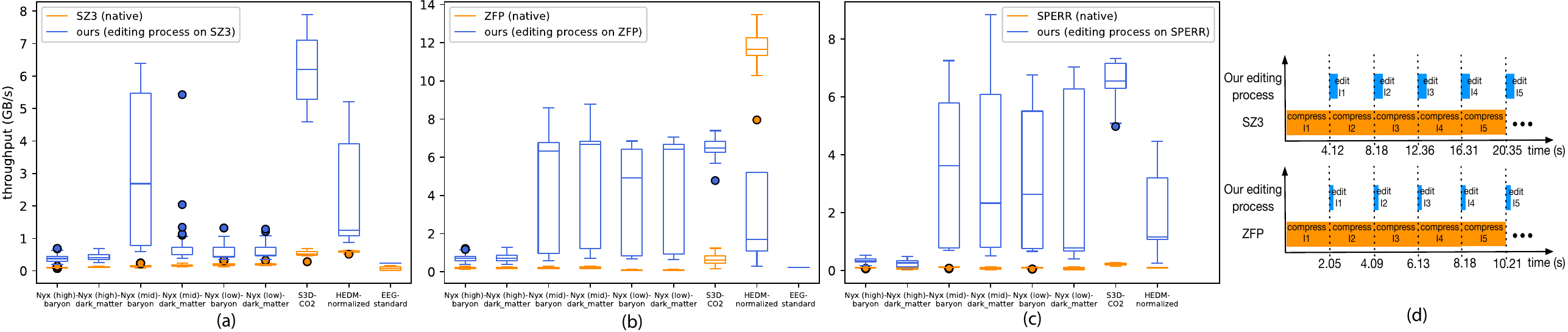}
    \caption{(a-c) Comparison of average throughput over different error bounds among our method, SZ3, ZFP, and SPERR. Our editing processes exclude the compression by base compressors. (d) Timelines of pipelined compression–editing workflow for multiple instances from Nyx (lowRes) by SZ3 and ZFP augmented by our method. Our editing process only takes a small percentage of time compared to SZ3's or ZFP's compression process, which does not influence the overall time of the workflow.}
    \label{fig:throughput}
\end{figure*}

\textbf{Metrics}. We compare different methods by quantitative metrics such as \textit{compression ratio} that our method introduces to the base compressors (Table~\ref{tab:storage_overhead}), \textit{rate distortion} (Fig.~\ref{fig:rate_distortion}), and \textit{throughput} (Fig.~\ref{fig:throughput}). Specifically, we evaluate the accuracy in the frequency domain by spectral signal-to-noise ratio (SSNR)~\cite{penczek2002three,unser1987new}:
\begin{equation*}
    SSNR(\hat{X}_k,X_k)=10\log_{10}\left(\frac{\sum_k|X_k|^2}{\sum_k|X_k-\hat{X}_k|^2}\right).
\end{equation*}

\begin{figure*}[!ht]
    \centering
    \includegraphics[width=\textwidth]{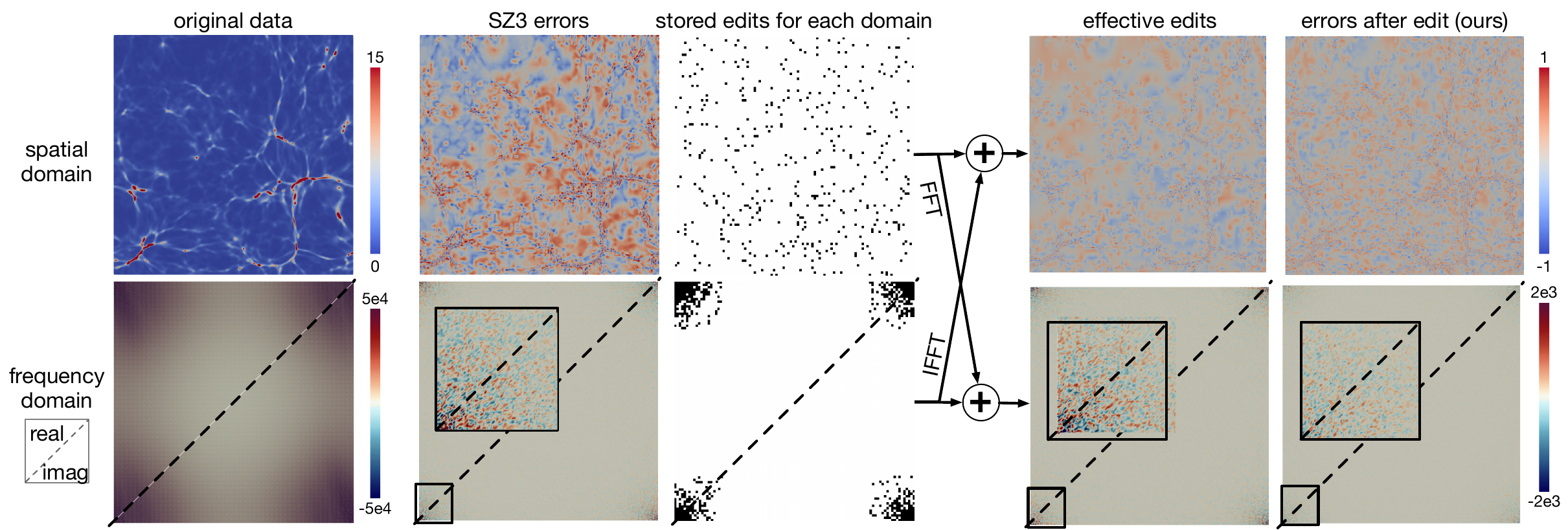}
    \caption{Visualizations on a 2D slice of our method applied to SZ3-compressed baryon density field in low-resolution Nyx data with $\epsilon=1$ and $\delta=2,000$. The third column displays the positions of spatial and frequency edits produced by our method, each modifying components in a single domain. There are 412 and 1706 active spatial and frequency edits, respectively, sparsely distributing. The total edits in a given domain shown in the fourth column are obtained by summing the two domain-specific edits, with forward or inverse FFT applied as needed to represent them in the same domain.}
    \label{fig:vis}
\end{figure*}

Higher SSNR implies smaller distortion in the frequency domain. We do not use peak signal-to-noise ratio (PSNR) for the frequency domain due to the energy conservation between the spatial and frequency domains by Parseval's theorem~\cite{hassanzadeh2022linear}, which implies that the mean squared error (MSE) is also preserved by FFT:
\begin{equation*}
    MSE(\hat{X}_k,X_k)=MSE(\hat{x}_n,x_n).
\end{equation*}
Thus, PSNR differs between the two domains only by a normalization factor, while SSNR provides a more meaningful measure of accuracy in the frequency domain. The accuracy of decompressed data in the spatial domain is measured by PSNR. We define the \textit{relative frequency error} (RFE) as
\begin{equation*}
    \mbox{RFE of component }l=\frac{|\mbox{frequency error }l|}{\max_k|\mbox{frequency component } k|}.
\end{equation*}
We also compare these compression algorithms qualitatively, including power spectrum (Fig.~\ref{fig:power_spectrum}) and visualization of decompressed spatial data (Fig.~\ref{fig:vis}).

\begin{table}[!th]
    \centering
    \caption{Benchmark datasets}
    \begin{tabular}{c|c|c|c|c}
        \toprule
        dataset & dim & size & attributes & precision \\ \hline
        Nyx (hiRes) & 3D & $2,048^3$ & \multirow{3}{*}{\makecell{baryon density,\\dark matter density}} & \multirow{3}{*}{single} \\ \cline{1-3}
        Nyx (midRes) & 3D & $1,024^3$ &  & \\ \cline{1-3}
        Nyx (lowRes) & 3D & $512^3$ &  & \\ \cline{1-5}
        S3D & 3D & $500^3$ & \makecell{CH4, O2, CO, CO2,\\H2O, N2} & double \\ \cline{1-5}
        HEDM & 2D & $2,048^2$ & normalized & double \\ \cline{1-5}
        EEG & 1D & 31,000 & standard & double \\
        \bottomrule
    \end{tabular}
    \label{tab:datasets}
\end{table}

\begin{table*}[!th]
    \centering
    \caption{Compression ratios of our method and base compressors. For Nyx and EEG datasets, $\epsilon(\%)=0.1$ and $\delta(\%)=0.1$; for S3D dataset, $\epsilon(\%)=0.1$ and $\delta(\%)=10^{-5}$; for HEDM dataset, $\epsilon(\%)=0.1$ and $\delta(\%)=10^{-7}$.}
    \begin{tabular}{c|c|c|c|c|c|c|c|c|c|c}
        \toprule
        \multirow{3}{*}{dataset} & \multirow{3}{*}{attributes} & \multicolumn{9}{c}{compression ratio} \\ \cline{3-11}
        & & \multicolumn{3}{c|}{SZ3} & \multicolumn{3}{c|}{ZFP} & \multicolumn{3}{c}{SPERR} \\ \cline{3-11}
        & & $\epsilon$ only & $\epsilon$ and $\delta$ & \textbf{our aug.} & $\epsilon$ only & $\epsilon$ and $\delta$ & \textbf{our aug.} & $\epsilon$ only & $\epsilon$ and $\delta$ & \textbf{our aug.} \\ \hline
        \multirow{2}{*}{\makecell{Nyx\\(hiRes)}} & baryon density & 44,307.1 & 218.8 & 34,957.7 & 406.0 & 50.6 & 405.8 & 16,834.6 & 2,159.1 & 13,427.3 \\ \cline{2-11}
        & dark matter density & 931.6 & 11.2 & 924.2 & 36.4 & 6.8 & 36.4 & 532.2 & 103.1 & 532.0 \\ \hline
        \multirow{2}{*}{\makecell{Nyx\\(midRes)}} & baryon density & 28,981.0 & 108.3 & 22,737.3 & 224.0 & 50.9 & 224.0 & 10,886.7 & 1,328.0 & 9,532.1 \\ \cline{2-11}
        & dark matter density & 645.0 & 11.0 & 636.6 & 19.5 & 5.7 & 19.5 & 377.5 & 96.9 & 377.4 \\ \hline
        \multirow{2}{*}{\makecell{Nyx\\(lowRes)}} & baryon density & 9,946.0 & 71.4 & 7,231.0 & 148.0 & 51.3 & 148.0 & 3,907.2 & 661.1 & 3,126.9 \\ \cline{2-11}
        & dark matter density & 334.9 & 11.3 & 327.8 & 12.4 & 5.7 & 12.4 & 210.4 & 102.9 & 210.4 \\ \hline
        S3D & CO2 & 2,822.6 & 100.9 & 2,514.3 & 95.1 & 35.4 & 95.1 & 1,864.0 & 700.3 & 1,864.4 \\ \hline
        HEDM & normalized & 4,632.5 & 3,190.0 & 3,813.4 & 628.9 & 567.2 & 628.9 & 993.9 & 256.5 & 993.9 \\ \hline
        EEG & standard & 13.7 & 11.5 & 13.7 & 1.8 & 1.2 & 1.8 & - & - & - \\
        \bottomrule
    \end{tabular}
    \label{tab:storage_overhead}
\end{table*}

\textbf{Datasets}. The datasets used in our evaluation, summarized in Table~\ref{tab:datasets}, span multiple scientific domains, spatial and temporal resolutions, and data modalities, and represents the diverse challenges faced by scientific data analysis. Nyx data is generated from cosmological simulations carried out with the Nyx code~\cite{almgren2013nyx}. S3D data comes from combustion simulations that resolve every detail of turbulent reactive flows at small scales, including fluid dynamics and detailed chemistry~\cite{sdrbench}. High-Energy Diffraction Microscopy (HEDM) data captures the intensities of diffracted rays from X-ray imaging of crystalline samples with high temporal resolution. Electroencephalography (EEG)~\cite{eeg} is a medical time-series database recording continuous brain activity.

\textbf{Platform}. We run experiments on the Perlmutter supercomputer at the National Energy Research Scientific Computing Center (NERSC)~\cite{perlmutter}. Each experiment used a single NVIDIA A100 GPU (40 GB HBM2 memory) running CUDA 12.4 or a single AMD EPYC 7763 CPU with 64 cores.

\subsection{Key observations}

\vspace{0.2cm}
\noindent\fbox{\begin{minipage}{24.5em}
\textbf{Observation 1}. The edits introduced by our method result in only a modest reduction in compression ratio (around 15\% for SZ3 and SPERR, and 0.001\% for ZFP) relative to the corresponding base compressors.
\end{minipage}}
\vspace{0.2cm}

Table~\ref{tab:storage_overhead} reports the compressed data sizes for three cases: (1) base compressors with only spatial errors bounded (native), (2) base compressors with both spatial and frequency errors bounded by tightening the user-specified spatial error bound (trial-and-error), and (3) our augmentation, which edits the results of case (1) to additionally bound frequency errors. For all datasets, the relative spatial error bound is fixed at $\epsilon(\%)=0.1$. The RFE bounds are selected such that, when using base compressors with $\epsilon(\%)=0.1$, the maximum frequency error of the reconstructed data are reduced by a factor of 100.

The compression ratio impact of our edits is minimal. Across all datasets, the trial-and-error approach substantially degrades the compression ratio, sometimes by orders of magnitude. In contrast, our method enforces frequency-domain bounds on top of the native compressor outputs while retaining compression ratios close to the original.

The effect of our edits varies by base compressor. For SZ3, the edits introduced by our method reduce the compression ratio by only about 10-20\%, which does not dominate the overall compressed size. For ZFP and SPERR, the effect is smaller and even negligible for ZFP (on the order of $10^{-5}$). This difference arises because SZ3, as a prediction-based compressor, estimates each data point from local neighbors without leveraging global correlations, making it less effective at preserving frequency content—particularly in high-frequency ranges. In contrast, ZFP and SPERR, as transform-based methods, exploit correlations across a broader spatial extent, inherently retaining more frequency-domain structure.

The modest impact on compression ratio of our method can be attributed to the limited number of active edits in both the spatial and frequency domains, as illustrated in Fig.~\ref{fig:vis}. In the third column, we see that the non-zero (active) edits are sparsely distributed, which means most spatial and frequency components remain unchanged. However, when we visualize the difference before and after editing in only one domain, computed as the sum of edits from both domains with forward or inverse transforms applied as needed, the changes appear across all components in that domain (see the fourth column). This observation also clarifies why we maintain two separate sets of edits, one for each domain.

\vspace{0.2cm}
\noindent\fbox{\begin{minipage}{24.5em}
\textbf{Observation 2}. For the same bitrate (or storage size), our method achieves higher frequency-domain accuracy and comparable spatial-domain accuracy compared with the baselines.
\end{minipage}}
\vspace{0.2cm}

\begin{wrapfigure}{LH}{0.4\linewidth}
    \centering
    \includegraphics[width=\linewidth]{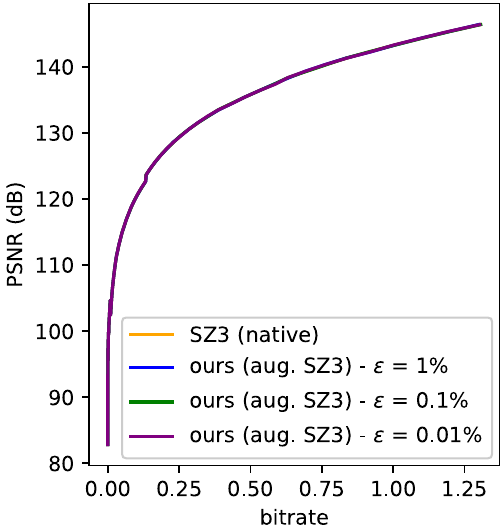}
    \caption{PSNR vs. bitrate of our method and SZ3 on spatial domain of Nyx (hiRes) data's baryon density. We edit on the results of base compressors with spatial error bounds $\epsilon(\%)=1$, $0.1$, and $0.01$.}
    \label{fig:spatial_rate_distortion}
\end{wrapfigure}

We plot SSNR versus bitrate for all methods in Fig.~\ref{fig:rate_distortion}, with our method applied to the results of base compressors at $\epsilon(\%)=0.1$. Our method consistently achieves higher SSNR at the same bitrate. For PSNR versus bitrate in the spatial domain (Fig.~\ref{fig:spatial_rate_distortion}), we observe that our method does not require additional storage to preserve spatial accuracy. Although it introduces additional storage for the edit data, the spatial errors are reduced after the editing process, resulting in only minor changes to the curves.

\vspace{0.2cm}
\noindent\fbox{\begin{minipage}{24.5em}
\textbf{Observation 3}. Our editing process is not the bottleneck of the overall running time.
\end{minipage}}
\vspace{0.2cm}

Figure~\ref{fig:throughput} (a-c) shows the throughputs of the base compressors and our editing process applied to them. The editing step is typically 2X-5X faster, indicating that it is not the bottleneck in overall runtime. The only exception is HEDM with ZFP, where the base compressor achieves throughput comparable to our editing process. This occurs because the HEDM dataset contains very few non-zero values, and the three base compressors employ different mechanisms to handle such data. The typical speed ranking is ZFP fastest, followed by SZ3, and then SPERR. This is because ZFP has a fast path for all-zero blocks: it performs a quick norm check on each block and then decides whether to emit a small or larger code. By contrast, SZ3 predicts and quantizes every data point without a block- or point-skipping mechanism. SPERR, on the other hand, applies a full multi-level transform and scans the entire dataset, making it generally slower than SZ3.

The runtime of our editing process is strongly correlated with the number of iterations in the alternating projection, as shown in Table~\ref{tab:num_iterations}. However, the iteration count is not necessarily proportional to the frequency error bound. Instead, it depends more on the initial position of the spatial error vector and on the geometry of the $s$- and $f$-cubes. A small frequency error bound (e.g., $10^{-4}\%$ or $10^{-5}\%$) corresponds to a small $f$-cube, which is often enclosed by the $s$-cube. In such cases, the projection typically terminates after the first iteration without introducing any non-zero (active) spatial edits. By contrast, when the $s$- and $f$-cubes partially overlap, more iterations are required, with the convergence depending on the angle between their projected boundaries.

\begin{table}[!th]
    \centering
    \caption{Number of iterations, active edits, and runnimg time in alternating projection of our editing process on SZ3 results of baryon density on Nyx(low) with $\epsilon(\%)=0.1$.}
    \begin{tabular}{c|c|c|c|c}
        \toprule
        $\delta(\%)$ & \# iters & \# act. spat. & \# act. freq. & time (ms) \\ \hline
        $10^{-2}$ & 53 & 445 & 2,325 & 628 \\ \hline
        $10^{-3}$ & 98 & 1770 & 323,789 & 946 \\ \hline
        $10^{-4}$ & 1 & 0 & 16,353,370 & 378 \\ \hline
        $10^{-5}$ & 1 & 0 & 53,950,920 & 396 \\
        \bottomrule
    \end{tabular}
    \label{tab:num_iterations}
\end{table}

\begin{figure}
    \centering
    \includegraphics[width=\linewidth]{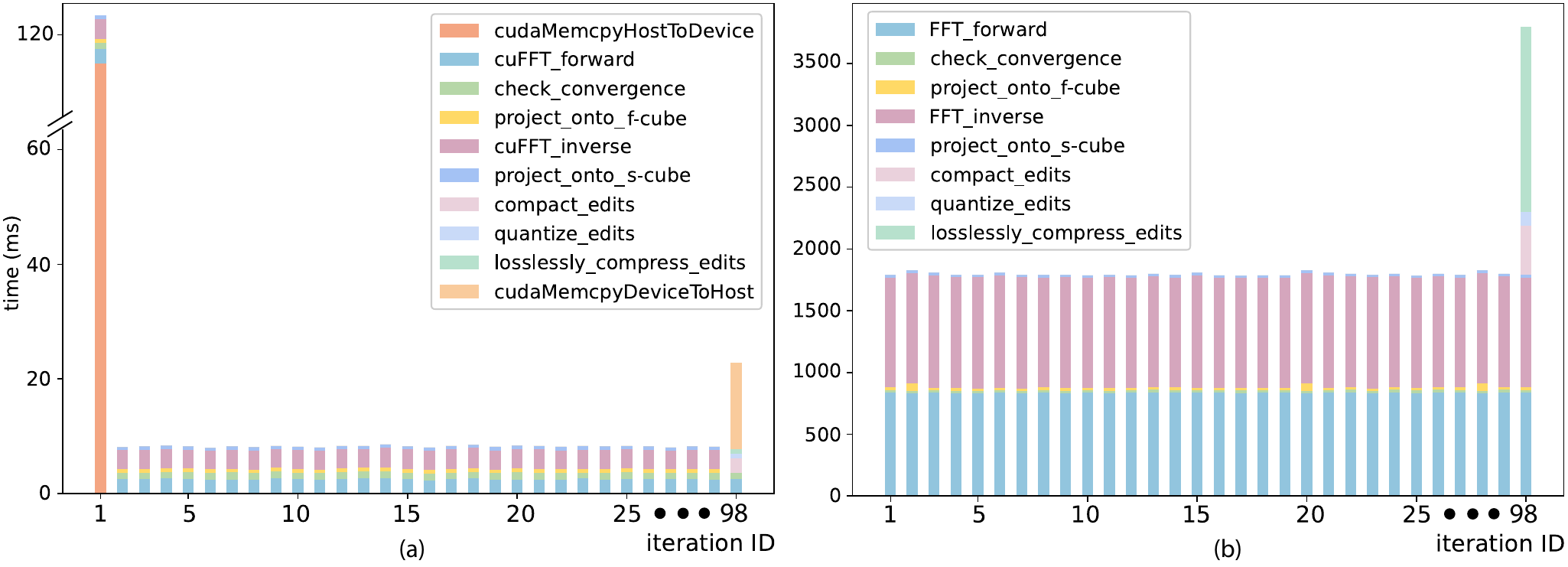}
    \caption{Timings of (a) GPU kernels and memory transfers and (b) CPU functions in our editing process applied to SZ3’s reconstruction of baryon density from the Nyx (lowRes) dataset, with $\epsilon(\%)=0.1$ and $\delta(\%)=0.001$. Several iterations are omitted due to the low variability in per-iteration timing.}
    \label{fig:iter_time}
\end{figure}

\begin{table*}[!th]
    \centering
    \caption{Performance of GPU kernels and CPU functions when editing SZ3's results with $\epsilon(\%)=0.1$ and $\delta(\%)=0.001$.  Metrics include execution time, effective memory bandwidth (BW), BW efficiency (BW Eff) relative to the hardware peak, floating-point performance (GFLOPS), FLOP efficiency (FLOP Eff), and arithmetic intensity (AI).}
    \begin{tabular}{c|c|c|c|c|c|c|c|c}
    \toprule
    kernel/function & platform & time (ms) & BW (GB/s) & BW Eff (\%) & GFLOPS & FLOP Eff (\%) & AI & Speedup \\ \hline
    \multirow{2}{*}{\makecell{forwardFFT\\(cuFFT / FFTW)}} & GPU & 2.53 & 423 & 26.4 & 7160 & 36.7 & 16.9 & \multirow{2}{*}{321} \\ \cline{2-8}
    & CPU & 813 & 1.32 & 0.66 & 22.3 & 0.29 & 16.9 & \\ \hline
    \multirow{2}{*}{\makecell{inverseFFT\\(cuFFT / FFTW)}} & GPU & 3.34 & 321 & 20.0 & 5420 & 27.8 & 16.9 & \multirow{2}{*}{272} \\ \cline{2-8}
    & CPU & 910 & 1.18 & 0.59 & 19.9 & 0.26 & 16.9 & \\ \hline
    \multirow{2}{*}{CheckConvergence} & GPU & 1.16 & 462 & 29 & 57.8 & 0.3 & 0.25 & \multirow{2}{*}{14.7} \\ \cline{2-8}
    & CPU & 17 & 31.6 & 16 & 7.89 & 0.1 & 0.25 & \\ \hline
    \multirow{2}{*}{ProjectOntoFCube} & GPU & 0.65 & 1070 & 67 & 2000 & 10.3 & 1.25 & \multirow{2}{*}{26.2} \\ \cline{2-8}
    & CPU & 17 & 127.4 & 64 & 159.2 & 2.0 & 1.25 & \\ \hline
    \multirow{2}{*}{ProjectOntoSCube} & GPU & 0.59 & 890 & 56 & 900 & 4.6 & 0.5 & \multirow{2}{*}{28.8} \\ \cline{2-8}
    & CPU & 17 & 31.6 & 16 & 7.89 & 0.1 & 0.25 & \\ \hline
    \multirow{2}{*}{CompactEdits} & GPU & 2.56 & 628 & 39 & 52 & 0.27 & 0.08 & \multirow{2}{*}{27.3} \\ \cline{2-8}
    & CPU & 70 & 23 & 11.5 & 1.92 & 0.025 & 0.08 & \\ \hline
    \multirow{2}{*}{QuantizeEdits} & GPU & 0.77 & 208 & 13 & 696 & 3.6 & 0.33 & \multirow{2}{*}{19.5} \\ \cline{2-8}
    & CPU & 15 & 107 & 53 & 35.7 & 0.46 & 0.33 & \\ \hline
    \multirow{2}{*}{LosslesslyCompressEdits} & GPU & 0.81 & 665 & 42 & 33 & 0.17 & 0.05 & \multirow{2}{*}{67.9} \\ \cline{2-8}
    & CPU & 55 & 9.8 & 4.9 & 0.49 & 0.006 & 0.05 & \\ \hline
    cudaMemcpyHostToDevice & GPU & 115 & 4.65 & 0.29 & 0 & 0 & 0 & - \\ \hline
    cudaMemcpyDeviceToHost & GPU & 15.2 & 35.3 & 2.2 & 0 & 0 & 0 & - \\ \hline
    \multirow{2}{*}{\textbf{end-to-end}} & GPU & \textbf{946} & \textbf{1100} & \textbf{69} & \textbf{4083} & \textbf{21} & \textbf{2.32} & \textbf{\multirow{2}{*}{184}} \\ \cline{2-8}
    & CPU & \textbf{174092} & \textbf{2.13} & \textbf{1.1} & \textbf{23.0} & \textbf{0.3} & \textbf{10.8} & \\
    \bottomrule
    \end{tabular}
    \label{tab:efficiency}
\end{table*}

Fig.~\ref{fig:throughput} (d) shows that our method does not increase the overall runtime in the pipelined compression–editing workflow for a sequence of data instances. When multiple instances are generated (e.g., under different parameters or at different time steps), compression and editing can be overlapped: the editing of instance $i$ is performed in parallel with the compression of instance $i+1$. As a result, the total runtime remains the same as that of a compression-only workflow.

We compare the GPU and CPU implementations of our correction algorithm at both the kernel/function level and end-to-end level, applied to SZ3’s reconstruction of baryon density from the Nyx (lowRes) dataset with $\epsilon(\%)=0.1$ and $\delta(\%)=0.001$ (Fig.~\ref{fig:iter_time}). This dataset was selected for its good CPU scalability, and the chosen error bounds allow more iterations, making per-iteration timing variability clearer. CPU functions are parallelized using OpenMP~\cite{openmp}. All timings exclude initial GPU warm-up and data loading overhead.

Among all GPU kernels, FFT/IFFT computations dominate, accounting for around 68.7\% of total kernel execution time. When the algorithm terminates after roughly 10 iterations, it exhibits memory-bound behavior, as memory transfers dominate runtime. For the case with 98 iterations, the algorithm becomes compute-bound, with kernel execution consuming 85.1\% of total GPU time and memory transfer only 14.9\%.

Table~\ref{tab:efficiency} reports the performance of GPU kernels and CPU functions for editing SZ3 results under the same parameters. Metrics include execution time (ms), effective memory bandwidth (GB/s) and its efficiency relative to the hardware peak, floating-point performance (GFLOPS) and FLOP efficiency, and arithmetic intensity (AI, FLOPs per byte accessed). For GPU kernels, the peak performance is 19.5 TFLOPS (FP32) with 1.6 TB/s HBM2 bandwidth; the CPU has a 7.8 TFLOPS FP32 peak and 200 GB/s memory bandwidth.

The GPU implementation provides substantial speedups over the CPU baseline, achieving 184X end-to-end acceleration and reducing total execution time from 174 seconds to under 1 second. Individual kernel/function speedups range from 14.7X to 321X. Analysis of AI and efficiency metrics shows that FFT kernels have moderate AI (around 17 FLOPs/byte) and high FLOP efficiency (around 30–37\%), reflecting a mix of compute-bound and memory-bound behavior. In contrast, most data-processing kernels, such as \texttt{CompactEdits} and \texttt{LosslesslyCompressEdits}, exhibit very low AI ($<1$ FLOP/byte) and low FLOP efficiency, indicating they are primarily memory-bound.

\begin{figure*}
    \centering
    \includegraphics[width=\textwidth]{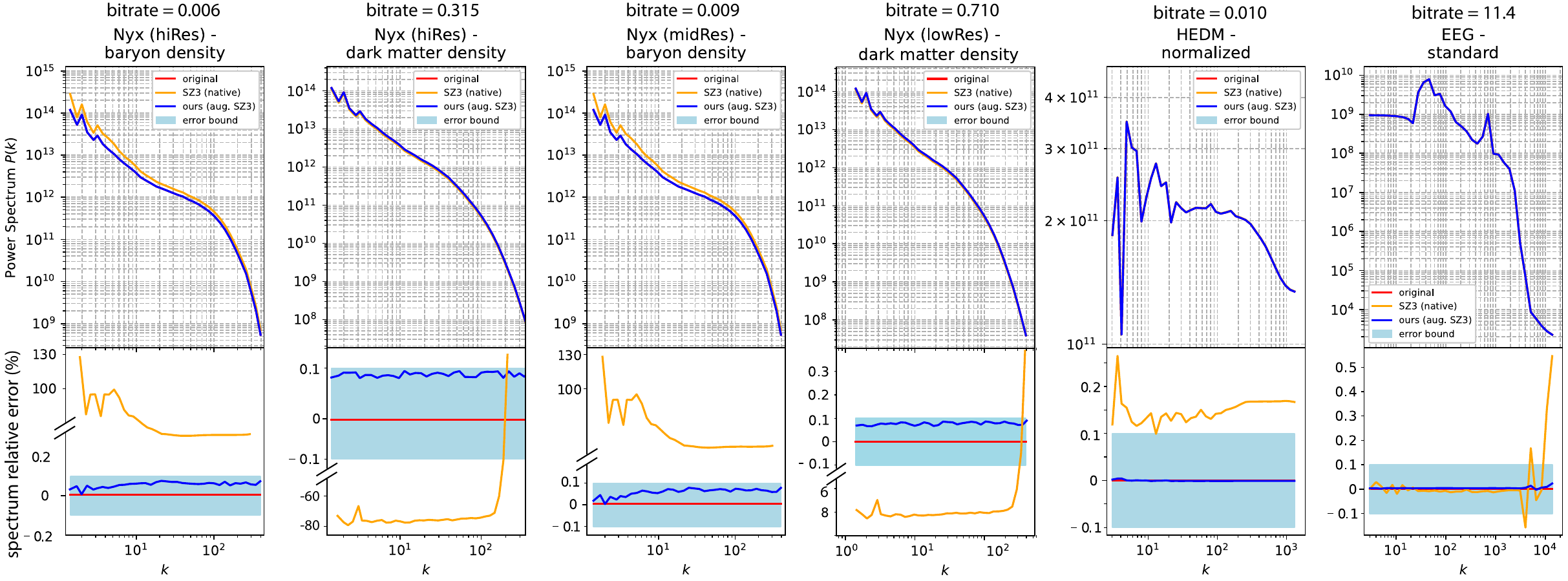}
    \caption{Power spectra and ratio of SZ3 and SZ3 augmented by our method, where ratio is the elementwise ratio of reconstructed power spectrum to the original one. We set the relative error bound of the power spectrum component to be 0.1\% for our method, as shown by the blue ribbon.}
    \label{fig:power_spectrum}
\end{figure*}

\vspace{0.2cm}
\noindent\fbox{\begin{minipage}{24.5em}
\textbf{Observation 4}. Our method preserves the power spectrum better under the same bitrate.
\end{minipage}}
\vspace{0.2cm}

Our method can preserve the power spectrum by assigning different pointwise relative error bounds to individual frequency components. In the upper row of Fig.~\ref{fig:power_spectrum}, we present our reconstructed power spectrum augmenting SZ3 where the relative error bound for each power spectrum component is set to 0.1\%. For comparison, the SZ3 results are shown under the same bitrate. The lower row plots the relative error of power spectrum, $\frac{\hat{P}(k)-P(k)}{P(k)}$, where $\hat{P}(k)$ is the reconstructed power spectrum obtained from our method or SZ3, and $P(k)$ is the original power spectrum. It is evident that our method consistently keeps the reconstructed spectrum within the specified error bound, whereas SZ3 occasionally exceeds it.

\section{Conclusion and future work}

We presented FFCz, a GPU-accelerated correction algorithm applicable to arbitrary base compressors, that guarantees data fidelity in both spatial and frequency domains, addressing a gap left by existing methods that preserve only spatial information. By transforming frequency-domain constraints into the spatial domain, FFCz represents the feasible region as the intersection of $s$- and $f$-cubes and enforces these constraints through iterative alternating projections. The method integrates seamlessly with compressors such as SZ3, ZFP, and SPERR, ensuring that their outputs satisfy user-specified dual-domain error bounds with small storage and runtime overhead. Our experiments on datasets from cosmology, X-ray diffraction, and combustion confirm that the proposed method preserves critical features essential for downstream analysis, enabling reliable dual-domain data interpretation.

Despite its strong performance, several limitations suggest directions for future work. First, the iterative projection algorithm can require many iterations to converge, particularly when the base compressor produces spatially irregular errors. Developing a direct or hybrid projection scheme could improve convergence predictability. Second, our method needs user-specified spatial and frequency error bounds, which are inherently coupled. Automated or data-driven approaches to jointly tune these bounds could improve usability and efficiency. Third, the post-hoc integration of our algorithm into existing compressors introduces some computational and memory overhead. Embedding feasibility enforcement directly into the compression pipeline could yield faster and more compact implementations. Finally, while demonstrated with Fourier-domain constraints, the approach naturally extends to other transforms (e.g., wavelets or spherical harmonics) and could be generalized to enforce fidelity across multiple domains—spatial, spectral, temporal, or multiresolution, broadening its applicability to diverse scientific workflows.

\section*{Acknowledgement}
The material was supported by the U.S. Department of
Energy, Office of Science, Advanced Scientific Computing Research
(ASCR), under contracts DE-AC02-06CH11357 and DE-SC0025677.
This research used resources of the National Energy Research Scientific Computing Center (NERSC), a Department of Energy User Facility using NERSC award DDR-ERCAP 0034457.

\bibliographystyle{IEEEtran}
\bibliography{references}

\end{document}